\documentclass{article}
\usepackage{graphicx} 
\usepackage[utf8]{inputenc}
\usepackage[english]{babel}
\usepackage[
backend=biber,
style=nature,
sorting=none,
]{biblatex}
\usepackage{authblk} 
\usepackage{amsmath}
\usepackage[left=1in,
  right=1in,
  top=1in,
  bottom=1in]{geometry} 
  
\usepackage{setspace}
\usepackage{xcolor}
\usepackage{url}

\graphicspath{ {./figs/} }

\usepackage{hyperref}
\hypersetup{backref=true,       
    pagebackref=true,               
    hyperindex=true,                
    colorlinks=true,                
    breaklinks=true,                
    urlcolor= black,                
    linkcolor= blue,                
    bookmarks=true,                 
    bookmarksopen=false,
    filecolor=black,
    citecolor=black,
    linkbordercolor=blue
}

\renewcommand{\d}[2]{\frac{d #1}{d #2}} 
 
\newcommand\blfootnote[1]{%
  \begingroup
  \renewcommand\thefootnote{}\footnote{#1}%
  \addtocounter{footnote}{-1}%
  \endgroup
}

\title{What Leads to Administrative Bloat? \\A Dynamic Model of Administrative Cost and Waste}

\author[a,b,1]{Vicky Chuqiao Yang}
\author[b]{Levi Grenier}

\date{}

\affil[a]{MIT Sloan School of Management, Massachusetts Institute of Technology, Cambridge MA, USA}
\affil[b]{Institute for Data, Systems, and Society, Massachusetts Institute of Technology, Cambridge MA, USA}

\begin{document}

\maketitle
\vspace{-1.5cm}

\blfootnote{\textsuperscript{1} Correspondence V.C.Y: vcyang@mit.edu}

\begin{abstract}
The functioning of complex systems depends on the coordination of diverse components, often supported by regulatory structures that incur costs. In human organizations, such costs manifest as administrative burden, which has been rising despite often reducing efficiency. Classic explanations point to bureaucrat self-interest or regulation, yet they do not explain variation across organizations or clarify how this burden can be reduced. Here, we develop a dynamical model of administrative growth that integrates known behavioral mechanisms of process creation, obsolescence, and removal. The model conceptualizes processes as developed for problem solving, but becoming obsolete as conditions change, while continuing to consume resources until actively pruned. This interplay generates two long-term outcomes: stable equilibrium or run-away growth. The threshold separating these outcomes is shaped by organizations' propensity to create new processes when faced with problems, and their propensity to prune obsolete ones in response to administrative burden. Importantly, their effects are asymmetric: sufficiently high creation propensity leads to bloat regardless of pruning propensity. Faster environmental change shifts this threshold, making bloat more likely. Simulations of interventions show that lasting reductions in administrative costs and waste require permanent shifts in priorities and investments in distinguishing obsolete from useful processes. Temporary efforts or indiscriminate cuts provide only short-lived relief, and counterintuitively, prioritizing direct production can increase waste. Our work highlights a general mechanism by which well-intentioned problem-solving can create self-reinforcing inefficiencies in complex systems, offering insights possibly generalizable to broader applications, such as legal, policy, and software systems where obsolete elements accumulate. 
\end{abstract}

\section*{Significance Statement}
Administrative costs have ballooned across industries—raising tuition, healthcare expenses, and employee burnout. Why does this burden persist, and how can it be reversed? We develop a dynamic model showing how well-intentioned problem-solving can, under certain conditions, lead to administrative bloat. Unlike classic explanations that portray administrative bloat as an inevitable tendency or regulatory imposition, our results highlight how differences in decision-making heuristics can tip systems toward either stability or run-away bloat. While faster environmental change makes bloat more likely, organizations can avoid it by permanently shifting priorities---tolerating some problems ad hoc and devoting sustained effort to pruning obsolete processes. Our results highlight a general systems mechanism: well-intentioned problem-solving can produce self-reinforcing inefficiencies. 

\section{Introduction}
The functioning of complex systems depends on the coordination of diverse components, often requiring dedicated regulatory functions to operate effectively. These regulatory functions incur costs, but are often essential for ensuring reliability and adaptability \cite{yang2024regulatory, yoon2023makes}. Yet across many domains of human society, these costs, manifested as administrative burden---rules, procedures, and the associated effort to administer them---have been growing persistently, often with counterproductive consequences. 

\begin{figure}[tb]
    \centering
    \includegraphics[width = 0.7\textwidth]{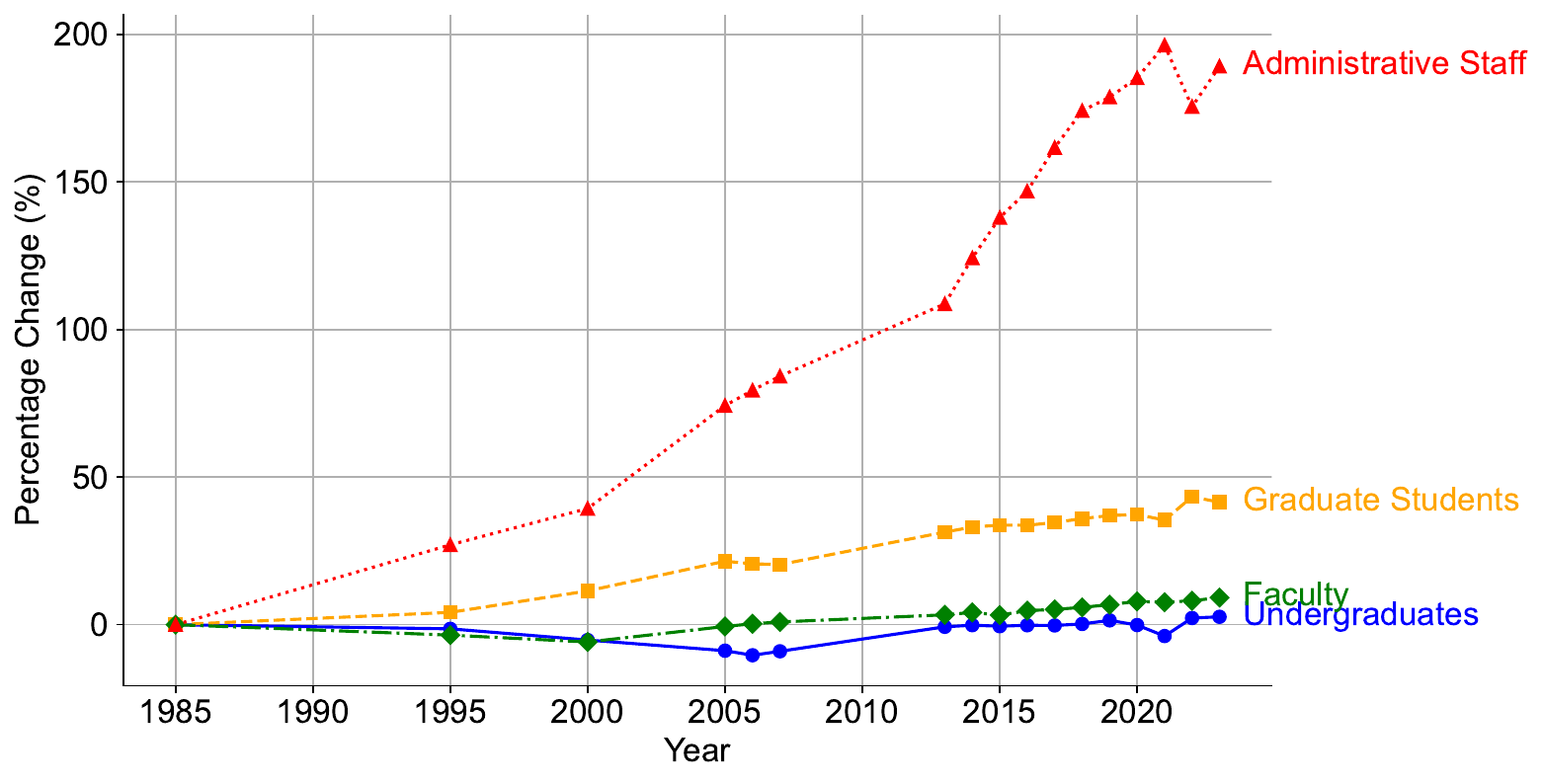}
    \caption{Percentage growth in administrative staff compared to the growth of undergraduates, graduate students, and faculty at MIT from 1985 to 2023.}
    \label{fig:mitData}
\end{figure}

Between 1983 and 2018, while total U.S. employment grew by 44\%, the number of managers and administrators more than doubled \cite{hamel2020humanocracy}. In higher education, administrative costs have risen much faster than instruction and research costs \cite{leslie1995rising}. One illustrative example is MIT. Faculty grew by only 9.2\% between 1985 and 2023, while administrative staff grew by 189\% (See Fig.~\ref{fig:mitData}). The U.S. spent approximately \$812 billion on healthcare administrative costs in 2017, representing 34.2\% of total healthcare expenditures \cite{himmelstein2020health}. Emergency department physicians spend  44\% of their time on data entry and only 28\% on direct patient care \cite{hill2013}. In renewable energy technologies, hardware costs have rapidly declined, but administration-related costs remain high, potentially creating a new bottleneck in the transition to green energy \cite{klemun2023mechanisms}. 

The rise in administrative costs is closely linked to growing employee frustration and declining efficiency. Consider a few examples: A university faculty member is required to produce a transcript from fifty-five years ago to prove they are qualified to teach a course they have already taught a dozen times \cite{vedder2020}. A service representative at a biotech firm must navigate fifteen different applications to perform basic tasks---the result of the IT manager continuously adding new ``efficiency'' software \cite{sutton2024friction}. Studies of health care and government employees have confirmed that pointless administrative tasks contribute to employees' alienation, lower job satisfaction, burnout, and attrition \cite{thun2018study, dehart2005red, rao2017impact}. 

One long-standing account for administrative expansion attributes it to bureaucrats' empire-building tendencies---administrators hiring subordinates to amplify their influence, secure budget growth, and protect their positions \cite{niskanen2017bureaucracy, kaufman1976government, meyer2013limits, ginsberg2011fall}. Another prominent explanation emphasizes external regulatory demands—laws, accreditation and compliance requirements---that fuel administrative growth by obligating organizations to add layers of reporting, monitoring, and coordination staff \cite{johnson2020administrative, bozeman2000}. While these explanations highlight real pressures, they would also suggest that administrative bloat is an inevitable outcome. These explanations fail to account for notable exceptions where organizations have successfully reversed or contained administrative growth, and offer limited practical guidance for managing administrative costs. Although rising concern over administrative burden has led to many intervention efforts, most have experienced only short-term relief, with administrative burden eventually rebounding \cite{sutton2024friction, balzer2020lean}. A small number of organizations, however, represent notable exceptions. Some of the most dramatic cases include Haier, a Chinese appliance manufacturing company, which eliminated its entire 12,000 middle manager positions while sustaining growth and industry leadership \cite{kanter2018haier}; and General Electric’s jet engine plant in Durham, North Carolina, which has been operating with more than 300 technicians overseen by a single manager \cite{hamel2016excess}.



Addressing this gap requires a scientific framework that can trace the dynamics of administrative growth to first principles. We address this by proposing a dynamical model integrating behavioral mechanisms governing the creation, obsolescence, and removal of codified processes. March et al. \cite{march2000dynamics} analyzed a century of Stanford University faculty meeting records through qualitative and statistical methods, providing rare empirical evidence of these mechanisms. Their work begins from the premise that organizations operate under competing priorities for limited resources. Rule creation and removal are two such competing activities. March et al.'s empirical analysis suggests creation is driven by attempts to address unsolved problems, often coming from the external environment, while removal depends less on the external environment and more on internal attention—for instance, complaints about excessive administrative burden \cite{march2000dynamics}. Crucially, the usefulness of rules depends on their alignment with the current environment: rules function as a system's memory, valuable when aligned with the environment but obsolete once misaligned.

While March et al.~identified three key behavioral components contributing to administrative cost and waste, these factors were studied separately, and there is no quantitative, system-level account of how these factors interact dynamically to give rise to the level of administrative cost and waste in an organization. Lacking this system-level integration, these insights do not specify the conditions under which administrative costs will grow unchecked, or when this trajectory can be reversed.

In this paper, we formalize a quantitative, dynamic model, integrating the three mechanisms of process creation, obsolescence, and removal for a system under resource constraints. It identifies the conditions under which well-intentioned problem-solving can lead to self-reinforcing accumulation of administrative waste, and when it can be kept in check, and enables an exploration of various intervention strategies. We test several strategies in simulation for how effective they are at reducing administrative cost and waste, including permanent and temporary changes of priorities, indiscriminate removal of processes, and prioritizing the ``real work.''

\section{Summary of the Mathematical Model}
\begin{figure*}[htb]
    \centering
    \includegraphics[width =0.85\textwidth]{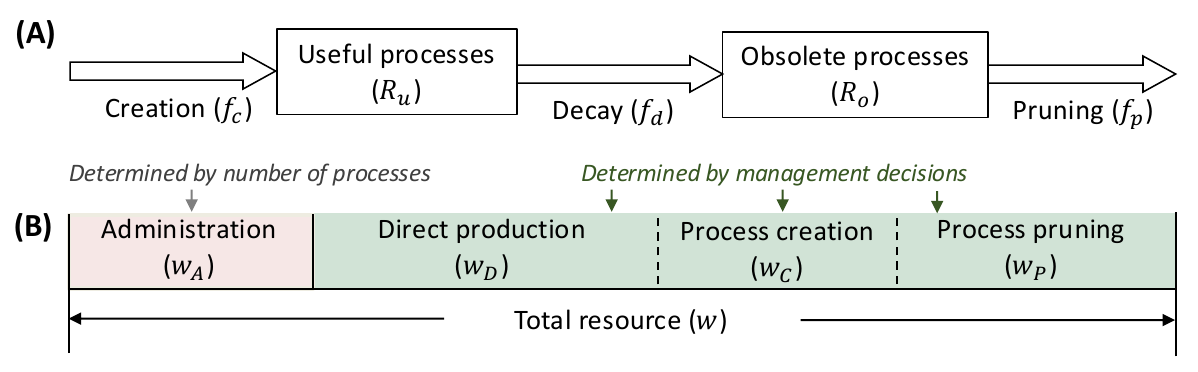}
    \caption{Conceptual diagrams illustrating model components. (A) Illustration of the key dynamics in the model relating useful and obsolete processes. (B) Organizations are constrained by fixed resources, which are allocated to four categories.}
    \label{fig:schematic_diagram}
\end{figure*}

The model considers a top-down managed organization---a structure that spans sectors from corporations to universities to government agencies. The organization faces problems that evolve over time as the external environment changes. In response, it can create codified processes to address unsolved problems. Examples of codified processes include employee reporting forms, approval requirements, and compliance training. However, as conditions shift, some problems change or become irrelevant, rendering the corresponding processes obsolete. These obsolete processes, although no longer useful, continue to be administered and consume resources, until some resources are invested in removing them. Figure~\ref{fig:schematic_diagram}(A) illustrates the relationship between the two kinds of processes. The numbers of useful and obsolete processes are denoted as $R_u$ and $R_o$, respectively. For tractability, we make the optimistic assumption that all processes are useful when created, and all removed processes are obsolete. 

To capture these dynamics parsimoniously, we choose to formulate a continuous-time differential equation model. This modeling framework captures systems of interacting and time-varying variables, accounting for any feedback loops and delays \cite{sterman2002system}, and enables the analysis of rich nonlinear behavior, including identifying multiple equilibria and tipping points \cite{strogatz2024nonlinear}. Models utilizing this method have been fruitful in the study of organizational dynamics, including the accumulation and decay of organizations' capability \cite{repenning2002capability}, and identifying the consequences of short-sighted managerial decisions \cite{rahmandad2018making}.  Because the mechanisms of interest do not depend on heterogeneity among individual agents or processes, we adopt this low-dimensional deterministic formulation rather than an agent-based model. 

These process creation, decay, and pruning dynamics can be mathematically described with a set of two differential equations, 
\begin{equation} \label{eq:stock_flow_odes}
    \d{R_u(t)}{t} = f_c (t) - f_d (t) \;,
    \d{R_o (t)}{t} = f_d (t) - f_p (t) \;.
\end{equation}
where $f_c(t)$, $f_p(t)$, and $f_d(t)$ correspond to the flows of processes created, pruned, and decayed from useful to obsolete, respectively. We denote the average time for a process to become obsolete as $T_d$, largely determined by the rate of environmental change. The flow of process decay is then $f_d(t) = R_u(t)/T_d$. The remainder of the model focuses on deriving the flows $f_c (t)$ and $f_p(t)$. We give a summary of the derivation here, with mathematical details provided in Materials and Methods. 

We consider the organization to be constrained by finite resources ($w$), representing the total work hours available from employees. Some of these resources are absorbed by the administration of existing processes (both useful and obsolete). The organization decides how to allocate the remaining resources among three competing activities: creating new processes to address unsolved problems, identifying and removing obsolete processes to reduce administrative burden, and engaging in direct production. Figure~\ref{fig:schematic_diagram}(B) depicts the different streams of resource allocation.  

At any given time, we consider the organization faces a fixed number of problems, though the specific problems may vary over time. Useful processes solve or prevent some problems, reducing the number of unresolved ones. Following March et al.\cite{march2000dynamics}’s findings, we consider that managers allocate resources to create new processes in response to unresolved problems. This tendency is captured by the \textbf{creation propensity} parameter, $\gamma_c$. A higher $\gamma_c$ means managers are more likely to create a process for each unresolved problem, while a lower value implies greater tolerance for problems or reliance on ad hoc solutions. Similarly, managers respond to administrative burdens—often raised through complaints—by allocating resources to remove obsolete processes. This response is represented by the \textbf{pruning propensity} parameter, $\gamma_p$, where higher values indicate greater effort devoted to pruning at a given level of administrative burden. Finally, managers must also allocate resources to direct production, whose priority is denoted by $\tilde d$. The flows of process creation and removal, $f_c(t)$ and $f_p(t)$, are proportional to resources allocated to process creation and removal. 

The model outputs the dynamics of two key outcome variables: \textbf{administrative cost}, the share of organizational resources taken up by process administration, and \textbf{administrative waste}, the share of resources taken up by administration of obsolete processes. 

\section{Results}
\begin{figure*}[htb]
    \centering
    \includegraphics[width = \textwidth]{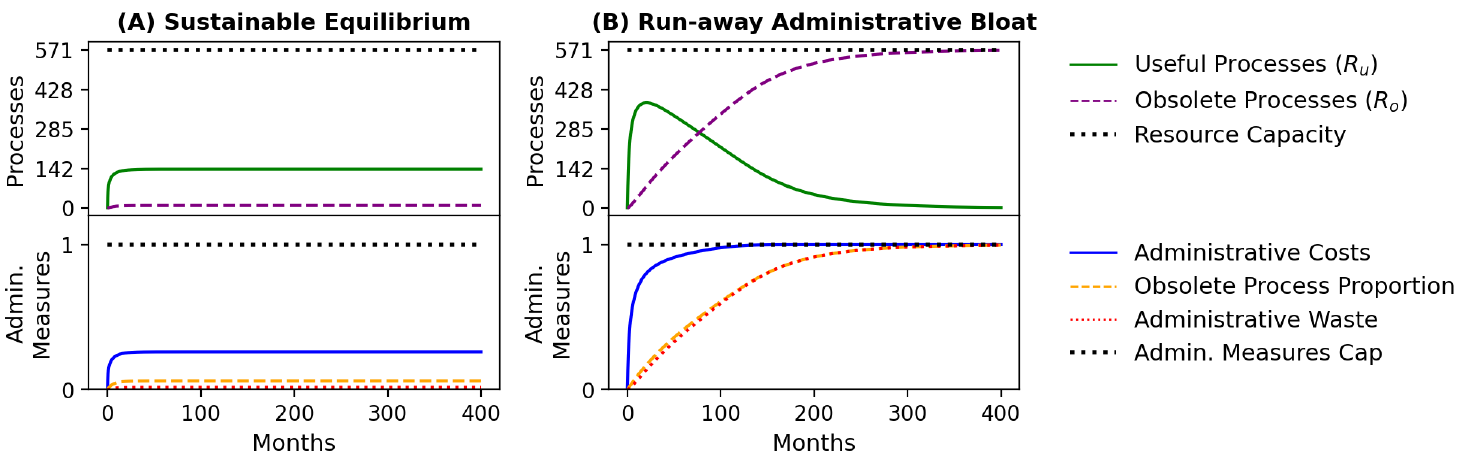}
    \caption{Model predicts two types of possible long-term outcomes. (A) Example of reaching a sustainable level of administrative costs and waste, below resource cap. (B) Example of run-away administrative bloat---administrative costs and waste grow to the maximum allowed by resources.}
    \label{fig:transient}
\end{figure*}

We simulate a system starting from the initial condition of one useful process and no obsolete ones. The model produces two distinct long-term outcomes. In the Sustainable Equilibrium regime (Fig.~\ref{fig:transient}(A)), useful and obsolete processes both stabilize at levels well below the resource capacity. Administrative waste and cost also plateau at intermediate values, never reaching the maximum possible of one. In the contrasting regime (Fig.~\ref{fig:transient}(B)), obsolete processes expand to resource cap, driving waste and cost to their theoretical limits—a run-away outcome that we term Administrative Bloat. In practice, we rarely see an organization exhaust its resources on administration. Systems drifting toward this trajectory are more likely to fail before reaching this limit for proximate reasons such as declining revenues, competitive pressures, and workforce attrition. Administrative bloat marks a boundary that organizations must avoid to remain viable. Model parameters, default values, and rationales are provided in Materials and Methods.

\begin{figure}[htb]
    \centering
    \includegraphics[width = 0.8\textwidth]{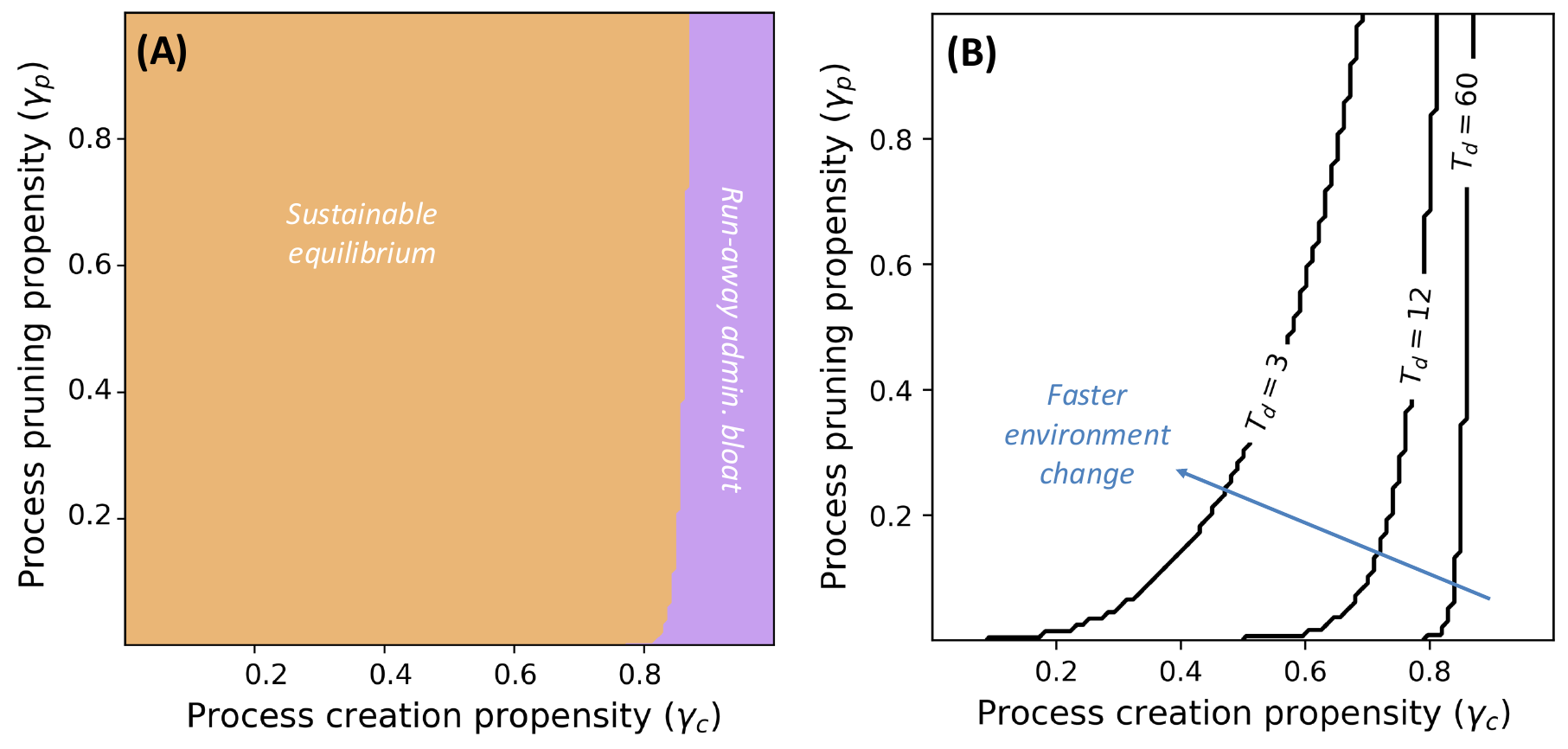}
    \caption{(A) Parameter regions for creation and pruning propensity leading to sustainable equilibrium and run-away administrative bloat outcomes. (B) The frontier that distinguishes the two kinds of outcomes for three values of process obsolescent time ($T_d$, in months), where shorter obsolescence time denotes faster environmental change. 
    }

    \label{fig:phase_plot}
\end{figure}

We delineate the conditions for each outcome by analyzing the equilibria of the differential equation system and their stability (see Supplementary Information for methods). Two key parameters govern the long-term outcome: creation propensity ($\gamma_c$) and pruning propensity ($\gamma_p$). The phase diagram in Fig.~\ref{fig:phase_plot}(A) shows a frontier separating two regimes. To the left of this threshold, administrative cost and waste stabilize at sustainable levels (Fig.~\ref{fig:transient}(A)); to the right, they exhibit run-away growth toward capacity (Fig.~\ref{fig:transient}(B)). Notably, when $\gamma_c$ exceeds a critical value, bloat occurs regardless of $\gamma_p$, indicating an asymmetry: high rates of process creation can entrench administrative bloat that pruning alone cannot reverse. In the Supplementary Information, we perform a robustness test and show that the two outcome regimes also arise when organizations do not follow behavioral heuristics but instead seek to maximize performance. In this case, bloat arises when the optimization horizon is short-sighted.

The intuition for these two regimes follows from the model’s interacting feedback loops. Several balancing loops counteract the growth of administrative cost. For example, creating useful processes reduces unresolved problems and suppresses further process creation, and rising administrative cost increases pruning priority, which in turn reduces that cost. Counteracting these balancing feedback loops is a reinforcing feedback loop: as the total number of processes grows, it reduces the free organizational resources that could be allocated to process pruning. This reduction in pruning capacity allows more obsolete processes to accumulate, further increasing administrative cost. The interplay between these balancing and reinforcing loops determines the system’s dynamics and helps explain the asymmetry between creation and pruning propensities in Fig.~\ref{fig:phase_plot}(A). When process creation propensity ($\gamma_c$) is low, balancing feedbacks dominate, yielding a stable, sustainable equilibrium. As $\gamma_c$ increases, the reinforcing loop strengthens. After $\gamma_c$ crosses a critical threshold, a tipping point occurs---administrative accumulation becomes self-sustaining, leading to run-away administrative bloat (see Supplementary Information for a more mathematical discussion of this tipping point as a saddle-node bifurcation). Pruning may operate as a delayed corrective mechanism that struggles to keep pace with sustained high rates of process creation, which reduces the capacity available for pruning.

\begin{figure*}
   \centering    
    \includegraphics[width = 0.95\textwidth]{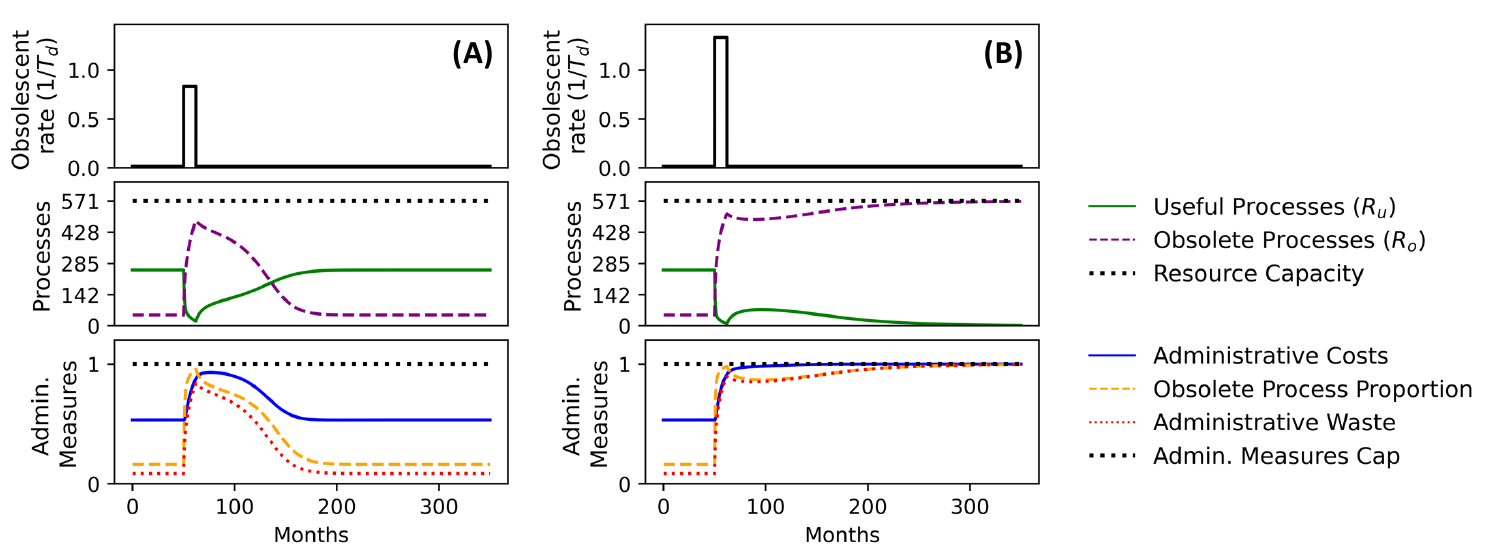}
        \caption{Effect of environmental shocks on administrative cost and waste. (A) Shocks below a critical threshold allow recovery to the original equilibrium. (B) Shocks above the threshold trigger a run-away cycle of administrative bloat. The system begins in equilibrium. The shock occurs at 50 months and lasts for 12 months.}
 \label{fig:shock}
\end{figure*}

An important parameter shaping the threshold is process obsolescence time ($T_d$), which captures the pace of environmental change. As shown in Fig.~\ref{fig:phase_plot}(B), faster change (smaller $T_d$) shifts the boundary leftward, making bloat more likely. Heuristics that yield sustainable outcomes in stable environments may therefore generate administrative bloat under rapid change. Faster change produces more obsolete processes, fueling more unresolved problems and further process creation. Achieving sustainability under these conditions requires lowering creation propensity---tolerating unresolved problems or addressing them ad hoc---and/or raising pruning propensity to proactively remove outdated processes. A temporary environmental shock can also trigger administrative bloat. Fig.~\ref{fig:shock} shows a system, starting at a sustainable equilibrium, experiencing a shock increase in process obsolescence rate (faster environmental change) that lasts 12 months. If the magnitude of the shock is smaller than a critical threshold, the organization eventually recovers to the original equilibrium after the shock subsides. If the shock's magnitude exceeds a critical threshold, it leads the organization to the run-away bloat equilibrium. 


\begin{figure*}[htb]
    \centering
    \includegraphics[width = 0.9\textwidth]{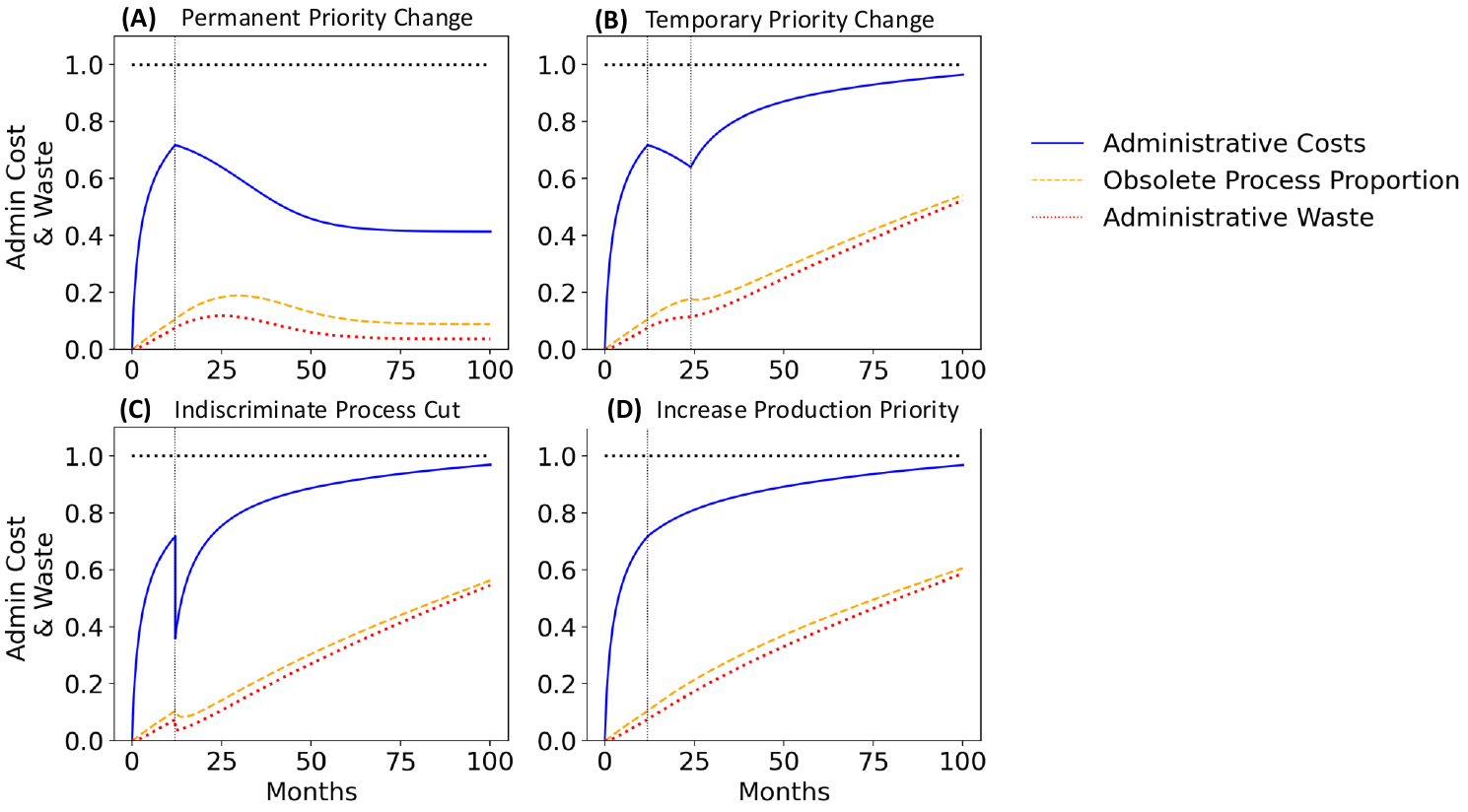}
     \caption{Effect of various interventions on administrative costs and wastes. The baseline (no intervention) condition is shown in Fig.~\ref{fig:transient}(B). (A) Permanent reduction in creation propensity ($\gamma_c$) and increase in pruning propensity ($\gamma_p$). (B) A stronger shift than (A), but the change in priorities is temporary. (C) One-time removal of processes, regardless of usefulness. (D) Increase direct production priority.}

       \label{fig:interventions}
\end{figure*}

Once an organization begins to drift toward administrative bloat, how can this trajectory be reversed? To address this question, we use the scenario in Fig.~\ref{fig:transient}(B) as the baseline, where bloat arises in the absence of intervention, and test several strategies introduced at month 12. The first strategy involves permanent adjustment of propensities: reducing the process reducing creation propensity ($\gamma_c$) from 0.9 to 0.7 and increasing pruning propensity ($\gamma_p$) from 0.5 to 0.7 for the rest of the simulation. As shown in Fig.~\ref{fig:interventions}(A), this shift stabilizes the system at lower levels of cost and waste, suggesting that permanent change in decision-making heuristics can counteract bloat. A second strategy is a temporary, but intensive intervention, reducing the creation propensity ($\gamma_c$) from 0.9 to 0.1 and increasing the propensity to prune a process ($\gamma_p$) from 0.5 to 0.9 for one year. While it produces short-term improvements (Fig.~\ref{fig:interventions}(B)), once the intervention ends, the system reverts to its prior trajectory, showing that temporary priority shifts do not substitute for lasting changes. A third strategy is a one-time, indiscriminate reduction in processes---removing half of the processes, regardless of their usefulness, as a broad organizational reset. While this intervention leads to a short-term decrease in administrative costs and waste, our model predicts that its effect to be temporary, and administrative costs and waste will bounce back in the long term (Fig.~\ref{fig:interventions}(C)).

\begin{figure*}[htb]
    \centering
    \includegraphics[width = 0.9\textwidth]{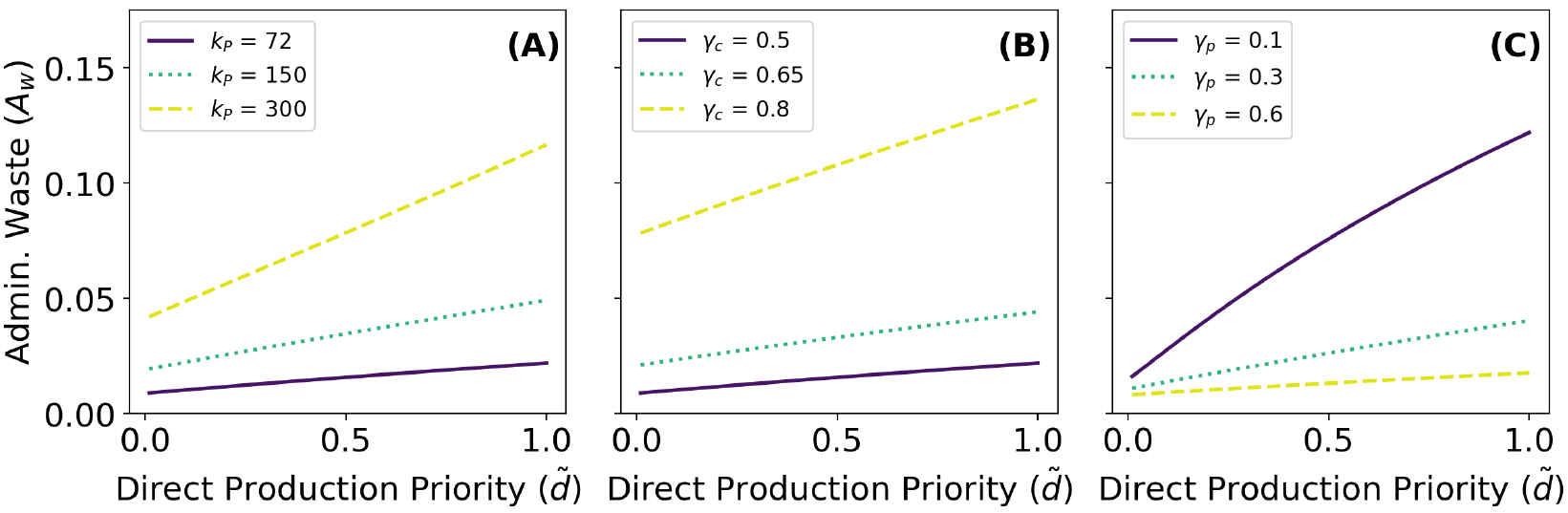}    
    \caption{Administrative waste tends to increase with direct production priority, across varying values of other parameters. (A) For various levels of pruning costs ($k_p$), (B) for various levels of process creation propensity ($\gamma_c$), (C) for various levels of process pruning propensity ($\gamma_p$).}
    \label{fig:waste_heatmap}
\end{figure*}

Finally, we consider increasing the priority of direct production, reasoning that prioritizing core outputs might constrain administrative overhead. This strategy not only fails to reverse the bloat trajectory (Fig.~\ref{fig:interventions}(D)). Surprisingly, the model predicts that higher production priority leads to more administrative waste. Fig.~ \ref{fig:waste_heatmap} shows the equilibrium solutions for administrative waste as a function of direct production priority (horizontal axis) while each of the three panels varies a parameter of the model: pruning cost, creation propensity, and pruning propensity. Across all three panels, we observe that as direct production priority increases, administrative waste tends to rise, or in the best-case scenario, stays the same. This occurs because directing attention to production suppresses both the creation of new useful processes and the pruning of obsolete ones. As useful processes inevitably decay, obsolete processes build up, driving waste upward even if the total number of processes falls.

\section{Discussion}

We develop a dynamical model that integrates process creation, obsolescence, and removal to identify conditions leading to administrative bloat. The system exhibits two regimes: a sustainable equilibrium and a run-away bloat state. Which regime emerges depends on heuristics for creating processes in response to problems and pruning them in response to administrative load. Faster environmental change shifts the threshold between these regimes, making bloat more likely. Thus, while accelerating change increases the risk of administrative bloat, the outcome remains contingent on organizational decision heuristics.

The results suggest that avoiding administrative bloat requires tolerating some non-critical problems or resolving them ad hoc, rather than codifying every potential issue into a process. This highlights the importance of discerning when structured solutions are necessary and when flexible responses suffice. For instance, codified processes are vital in high-stakes settings such as checklists in airplane maintenance, but unnecessary for low-stake contexts like office light bulbs---we simply replace them as they break. Effective decision-making lies in locating a problem on this “airplane–light bulb” spectrum, weighing administrative burden, potential error costs, and the expected lifetime of process usefulness.

We next examine whether intervention strategies can reverse the bloat trajectory. Our results highlight two key factors. The first is the continuity---only sustained change in creation and pruning propensities produces lasting reductions, whereas temporary changes, even intensive ones, have little long-term effect. The second is selectivity: distinguishing obsolete from useful processes is essential, since indiscriminate cuts merely delay the eventual bloat. Finally, shifting focus entirely to “real work,” while de-emphasizing both process creation and pruning, fails to prevent waste and can even exacerbate it. These dynamics clarify why some organizations successfully restrain administrative costs while others fall into escalating bloat.

Administrative bloat is often attributed to bureaucrats’ self-interest or external regulation---forces largely outside organizational control. Our model provides an alternative, endogenous explanation, showing that bloat can emerge as a result of well-intentioned problem solving, without bureaucratic self-interest or regulatory demands. Decision-making heuristics alone can generate divergent system trajectories. Our explanation does not aim to replace theories of exogenous pressure or misaligned incentives, but to complement them. What our model presents is a best-case scenario, or likely a lower bound on administrative cost. By formalizing these dynamics, the model also provides a foundation for future extensions that integrate external pressures and self-interest.

Our framework identifies a general mechanism through which efforts to adapt to a changing environment can, over time, generate self-reinforcing inefficiencies in complex systems. Although we focus on administrative bloat in organizations, the underlying dynamics are broadly applicable wherever rules, routines, or codes accumulate, become obsolete as conditions evolve, and impose a cost to the system to either carry them or remove them. Examples include legal and regulatory systems, where new statutes and compliance requirements are added more easily than they are repealed. In healthcare, billing and insurance codes proliferate while outdated ones remain. Software systems exhibit similar patterns: outdated legacy code persists widely, creating maintenance burdens and limiting new development. These parallels suggest that the conditions we identify for administrative bloat may represent a general risk pattern across domains.  Future work can extend this modeling framework to other settings and generate testable predictions about when adaptation leads to entrenched inefficiencies.

To focus on the core dynamics of process accumulation and pruning, our model is intentionally parsimonious. This simplicity enables clear identification of feedback mechanisms driving administrative bloat, though it necessarily omits several real-world complexities that represent opportunities for future research. One such simplification is that we omit considering the dependencies among processes. For instance, Institutional Review Board approval process typically requires prior completion of research-conduct training. Such dependencies can increase the cost of process removal, particularly in complex organizations where these relationships may be opaque. Future work could incorporate these interactions---for example, by applying a Design Structure Matrix approach \cite{eppinger2012design}---to investigate how modular process architectures might reduce pruning costs and mitigate administrative bloat. We also study a best-case scenario in which all newly created processes are initially useful and no external regulatory pressures exist. Relaxing them---for example, by allowing some processes to be obsolete from inception or by modeling regulatory mandates as exogenous drivers of problem space growth---would likely increase the risk of administrative bloat. Finally, our model assumes a fixed organizational resource pool to isolate the dynamics of process accumulation. In reality, resource availability often scales with organizational performance. An important extension would be to couple resource pool growth with organizational performance, allowing for analysis of how the risk of administrative bloat evolves as organizations expand in size.

Our dynamic model illustrates that well-meaning responses to solve problems can, through endogenous feedback, generate long-term inefficiency. We show that administrative bloat can thus arise even in the absence of ill-intentioned bureaucrats or regulatory excess, while pointing toward actionable interventions. By introducing a formal model grounded in empirical behavioral findings, we build a bridge between qualitative insights on how routines are created and removed, and the quantitative dynamics of resource allocation and process accumulation. We hope this work advances a system-level, predictive, and ultimately, prescriptive understanding of how systems can manage complexity and avoid the trap of ever-expanding administrative burden.

\subsection*{Code and data availability}
The code and data involved in this study are available at \\
\url{https://github.com/levigrenier/Yang-Grenier-Administrative-Bloat}. 


\subsection*{Acknowledgements}{V.C.Y and L.G. were partially supported by National Science Foundation award EF--2133863. V.C.Y and L.G. thank Nelson Repenning for helpful feedback on an earlier version of this manuscript; Hazhir Rahmandad and John Sterman for valuable discussions on model formulation.} 

\section*{Materials and Methods}
\subsection*{Detailed description of the mathematical model}

Here we present the detailed derivation for the flows of process creation, $f_c$ and process removal, $f_p$, following the model summary in the main text. 

\begin{figure}[htb]
    \centering
    \includegraphics[width =0.4\textwidth]{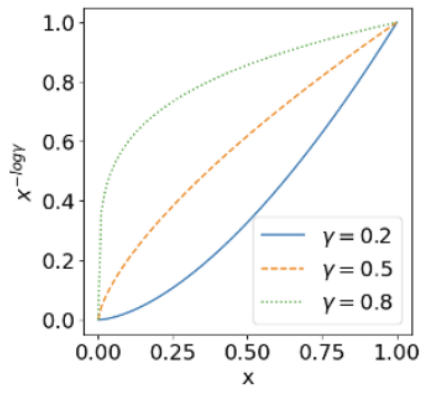}
    \caption{The function $y = x^{-\log(\gamma)}$, used to specify the organization priority for process creation and pruning for several $\gamma$ values.}
    \label{fig:gamma_function}
\end{figure}

The organization's resource constraint, $w$, equates to the aggregate available work hours. The organization allocates this resource (or effort) into four main activities: process administration ($w_A$), direct production ($w_D$), process creation ($w_C$), and process pruning ($w_P$), as visualized in Fig.~\ref{fig:schematic_diagram}(B). 

Process administration covers the efforts of employees engaged in process compliance and oversight, such as filling forms, responsible research conduct training, reimbursement approval processes, and executing checklists. Here, we consider that all processes are administered, and administration effort ($w_A$) is proportional to the number of processes, regardless of whether they are useful or obsolete, $w_A = k_A (R_o + R_u)$, where $k_A$ is the average effort to administer each process. Administration is prioritized in resource distribution, justified by the obligatory nature of many administrative tasks like prior authorizations, human resource processes, and performance reviews \cite{rao2017impact}. In the Supplementary Information, we relax this assumption and show our main conclusions hold under imperfect administration. The residual effort, $w - w_A$, is then divided among direct production, process creation, and process pruning. 

We then formulate the priorities of process creation, pruning, and direct production, denoted as $\tilde c$,  $\tilde p$, $\tilde d$, respectively. Following insights from March et al., we assume that pruning priority increases with administrative costs, reflecting the greater attention organizations pay to obsolete processes as complaints of burdens grows. One function satisfying this property is $\tilde p = (w_A/w)^{-\log \gamma_p}$, where $\gamma_p$, termed the \textbf{pruning propensity}, is a decision-making parameter ranging from 0 to 1. It reflects the organization's proactiveness to eliminate obsolete processes in response to rising administrative burden, with higher $\gamma_p$ values indicating more aggressive pruning. An illustration of this function for several $\gamma_p$ values is shown in Fig.~\ref{fig:gamma_function}. 

We assume the organization faces a constant overall problem volume. These problems may be solved (through useful processes) or unsolved. In either case, they become obsolete with timescale $T_d$ as external conditions change, and are replaced at the same rate by new problems so that the total problem volume remains constant. When a solved problem becomes obsolete, any process designed to address it also becomes obsolete. For example, digital security risks that once motivated some organizations to adopt periodic password-change policies have become less relevant as multi-factor authentication becomes widespread in the broader digital environment, rendering this problem and the associated processes obsolete. In the Supplementary Information, we relax this assumption by decoupling arrival and obsolescence rates, allowing size of the problem space to vary over time; the model displays similar qualitative patterns. 

Following March et al.'s findings, we assume that organizations assign greater priority to creating new processes when confronted with more unresolved problems. The volume of unresolved problems ($U$) should shrink with the number of useful processes, $R_u$, with diminishing returns. We operationalize this consideration with the decaying exponential function, $U = e^{-a R_u}$, where $a$ is a positive parameter denoting process effectiveness. Considering the priority of process creation increases with unsolved problem space monotonically, we use the same formulation as in the priority of pruning, $\tilde c = U^{-\log(\gamma_c)}$, represented by the same function in Fig.~\ref{fig:gamma_function}. The parameter $\gamma_c$, also between 0 and 1, is the \textbf{creation propensity} of the organization, reflecting the tendency of the organization to create processes in response to problems. This parameter reflects managerial discretion in preempting issues through process creation versus addressing them ad hoc.

We consider the priority of direct production, $\tilde d$, to be a decision parameter set by the organization's management. 

The remaining resource available, $w-w_A$, is allocated among creation, pruning, and direct production in proportion to the organization's priority for each. The proportion of remaining resources allocated towards creation is $c = \tilde c/(\tilde c + \tilde d + \tilde p)$. Similarly, that for pruning is $p = \tilde p/(\tilde c + \tilde d + \tilde p)$, and that for direct production effort is $d = \tilde d/(\tilde c + \tilde d + \tilde p)$.  

The flow of processes created in any time period is then proportional to the effort allocated to process creation,
\begin{equation}
    f_c = \frac{(w-w_A)c}{k_c}\;,
\end{equation}
where $k_c$ denotes resources required to create a process. 

Similar to process creation, the number of processes pruned should proportionally increase with pruning effort, $(w-w_A)p$. We further consider that when pruning processes, the organization must examine the efficacy of existing processes to distinguish if they are useful or obsolete. Consequently, the rate of process pruning is expected to increase with the proportion of obsolete processes. This is formulated as, 
\begin{equation}
    f_p = \frac{(w-w_A) p}{k_p} \frac{R_o}{R_u + R_o}\;,
\end{equation}
where $k_p$ is the baseline level of resources required to prune one process. 

Now that we have formulated all terms in Eq.~\ref{eq:stock_flow_odes}, we define the key outcome variables. \textbf{Administrative cost}, $A_c$, is the proportion of total organizational resources allocated to process administration, $A_c = w_A/w$. \textbf{Obsolete process proportion}, $r_o =  R_o/(R_o + R_u)$, is the proportion of obsolete processes in all processes. \textbf{Administrative waste}, $A_w$, the proportion of total organizational resources spent administering obsolete processes, $A_w = r_o A_c$. 


\subsection*{Model parameters and baseline simulation values}

\begin{table*}[h]
\caption{Model's parameter descriptions, units, and typical values in simulations}
\begin{center}
\begin{tabular}{p{1.5in} | p{0.15\linewidth} | p{2.5in} | p{0.08\linewidth}}
    \bf Parameter & \bf Unit & \bf Description & \bf Typical value\\
    \hline

$T_d$: Process obsolescence time & Month & Average time for a useful process to become obsolete. & 60\\ 
    \hline
$k_c$: Process creation cost & Hour/Process &  Hours needed to create a process. & 24 \\ 
    \hline
$k_p$: Process pruning cost & Hour/Process & Hours needed to remove a process. & 72 \\ 
    \hline
$k_A$: Process administration cost & Hour/Process/ Month & Total hours needed to administer a process per month.  & $15$\\
    \hline
$w$: Total resource & Hour/Month & Total work hours available. Assuming a 50-employee organization.  &8571\\
    \hline
$\tilde{d}$: Desired Production & Dimensionless & Desired proportion of available resources allocated towards direct production.   & 0.5\\ 
    \hline
 $\gamma_c$: Creation Propensity & Dimensionless & Parameter determining desired level of process creation in response to organization's problems. & 0.5 \\ 
    \hline
$\gamma_p$: Pruning Propensity & Dimensionless &Parameter determining desired level of process removal in response to administrative costs. & 0.5\\
    \hline
$a$: Process Efficacy & 1/Process & Coefficient determining the reduction of the problem space due to useful processes.  & 0.05\\ 
    \hline
\end{tabular}
\label{tab:param}
\end{center}

\end{table*}

Table~\ref{tab:param} summarizes the model parameters, their units, and default simulation values. Although precise values do not alter the model’s qualitative dynamics, we select magnitudes consistent with realistic organizational scales. Default values correspond to an organization of 50 employees, with processes becoming obsolete on average every 60 months (5 years). Creating a new process is assumed to require 24 person-hours (a three-person team working one day), while removing an existing process requires three times as much effort (72 person-hours), reflecting the greater difficulty of elimination. Each process also incurs an administrative cost of 15 hours per month ($k_A$), or roughly 18 minutes per employee per month. For parameters bounded between 0 and 1---desired production, creation propensity, and pruning propensity---we set default values at 0.5 and often varying them as independent variables in simulation. Process efficacy is set at 0.05, implying that approximately 14 processes are needed to address half of the problems. Additional information on parameter values specific to each figures' simulation is shown in Supplementary Information. 

\printbibliography

\end{document}